\documentclass[reprint,twocolumn,superscriptaddress,showkeys,
nofootinbib,notitlepage,amsmath,amssymb,floatfix]{revtex4-1}

\pdfoutput=1
\usepackage[margin=0.50in]{geometry}  
\usepackage{graphicx}
\usepackage{latexsym,amsmath,amssymb,lmodern,float,url}
\usepackage{natbib}
\usepackage{color}
\usepackage{multirow}
\usepackage{bm}
\usepackage{array}

\usepackage{mathtools}
\DeclarePairedDelimiter\bra{\langle}{\rvert}
\DeclarePairedDelimiter\ket{\lvert}{\rangle}
\DeclarePairedDelimiterX\braket[2]{\langle}{\rangle}{#1 \delimsize\vert #2}

\def\de{\delta^{\vphantom{1}}}
\def\bde{{\bar\delta}}

\def\QQ{{Q\bar Q}}
\def\kqQc{{\kappa_{qc}}}
\def\kqQs{{\kappa_{sc}}}
\def\kqQ{{\kappa_{qQ}}}

\def\dep{{\delta^{\prime\vphantom{1}}}}
\def\B{B^{\vphantom{1}}}
\def\bt{{\bar\theta}}
\def\hf{{\displaystyle{\frac 1 2}}}
\def\h3{{\displaystyle{\frac 3 2}}}
\def\pt{{\tilde{P}_{\frac{1}{2}}}}
\def\ptp{{\tilde{P}^{\prime\vphantom{1}}_{\frac{1}{2}}}}
\def\pr{{P_{\frac{3}{2}}}}
\def\kqQ{{\kappa_{qQ}}}
\def\kdQ{{\kappa_{\dep \bar{Q}}}}
\def\v8{{V_{8}}}
\def\vls{{V_{LS}}}
\def\vt{{V_{T}}}

\begin{document}
\title{Fine Structure of Pentaquark Multiplets in the Dynamical
Diquark Model}
\author{Jesse F. Giron}
\email{jgiron@lanl.gov}
\affiliation{Computational Physics Division (XCP), Los Alamos
National Laboratory, Los Alamos, NM 87545, USA}
\author{Richard F. Lebed}
\email{Richard.Lebed@asu.edu}
\affiliation{Department of Physics, Arizona State University, Tempe,
AZ 85287, USA}
\date{October, 2021}

\begin{abstract}
We apply the dynamical diquark model to predict the spectrum of
hidden-charm pentaquark states in both unflavored and
open-strange sectors.  Using only Hamiltonian parameters introduced
in the tetraquark $S$- and $P$-wave multiplets, the model naturally
produces the level spacing supported by the most recent LHCb results
for $P_c$ structures.  Furthermore, using model inputs obtained from
data of hidden-charm, open-strange tetraquarks ($Z_{cs}$), we predict
the spectrum of $P_{cs}$ states, including the recently observed 
$P_{cs}(4459)$.  We find all pentaquark candidates observed to date
to belong to the $1P$ multiplet and hence have positive parity.
\end{abstract}

\maketitle

\section{Introduction}

A recent burst of discoveries in the sector of heavy-quark exotic
hadrons has now pushed the number of observed candidates to over
50\@.  Multiple detailed reviews of these states have been published
in recent years~\cite{Lebed:2016hpi,Chen:2016qju,Hosaka:2016pey,
Esposito:2016noz,Guo:2017jvc,Ali:2017jda,Olsen:2017bmm,
Karliner:2017qhf,Yuan:2018inv,Liu:2019zoy,Brambilla:2019esw}, but
even such comprehensive reports have been unable to keep pace with
the spectacular rate of new experimental findings that have occurred
with remarkable regularity to the present day.

This paper focuses upon an analysis of the hidden-charm pentaquark
states, labeled $P_c$ and $P_{cs}$ in the nonstrange and strange
sectors, respectively.  The first candidates [$P_c(4450)$,
$P_c(4380)$] were observed in 2015 by the LHCb
Collaboration~\cite{Aaij:2015tga} as structures in the $J/\psi \, p$
spectrum in the decay $\Lambda_b \! \to \! J/\psi \, p K^-$.  Then
in 2019, LHCb resolved $P_c(4450)$ into two peaks, $P_c(4440)$ and
$P_c(4457)$, and observed a further $J/\psi \, p$ structure,
$P_c(4312)$~\cite{Aaij:2019vzc}.  Very recently, one more
$J/\psi \, p$ structure [$P_c(4437)$] has been observed by LHCb in
the decay $B_s^0 \! \to \! J/\psi \, p \bar p$~\cite{LHCb:2021chn},
but at a somewhat lower statistical significance ($> \! 3 \sigma$)
than the other $P_c$ states.  The only known strange candidate to
date, $P_{cs}(4459)$, was also very recently observed at LHCb as a
$J/\psi \, \Lambda$ structure in the decay $\Xi_b^- \! \to \! J/\psi
\, \Lambda K^-$~\cite{Aaij:2020gdg} (also at just over $3\sigma$).
The measured masses and widths of the states are collected in
Table~\ref{tab:PentaMasses}.
\begin{table}
\caption{Pentaquark candidate masses and as widths measured by the
LHCb Collaboration~\cite{Aaij:2015tga,Aaij:2019vzc,LHCb:2021chn,
Aaij:2020gdg}.}
\centering
\begin{tabular*}{\columnwidth}{@{\extracolsep{\fill}} c  c  c }
\hline\hline
State & Mass (MeV) & \ Width (MeV) \\
\hline
$P_c(4312)$ & $4311.9 \pm 0.7^{+6.8}_{-0.6}$ &
$9.8 \pm 2.7^{+3.7}_{-4.5}$ \\
$P_c(4337)$ & $4337^{+7 \ +2}_{-4 \ -2}$ &
$29^{+26 \ +14}_{-12 \ -14}$ \\
$P_c(4380)$ & $4380 \pm 8 \pm 29$ &
$205 \pm 18 \pm 86$ \\
$P_c(4440)$ & $4440.3 \pm 1.3^{+4.1}_{-4.7}$ &
$20.6 \pm 4.9^{+8.7}_{-10.1}$ \\
$P_c(4457)$ & $4457.3 \pm 0.6^{+4.1}_{-1.7}$ &
$6.4 \pm 2.0^{+5.7}_{-1.9}$ \\
$P_{cs}(4459)$ & $4458.8 \pm 2.9^{+4.7}_{-1.1}$ &
$17.3 \pm 6.5^{+8.0}_{-5.7}$ \\
\hline\hline
\end{tabular*}
\label{tab:PentaMasses}
\end{table}

Note that all of these measurements have been made by LHCb\@.  The
only independent evidence for $P_c$ states to date comes from the D0
Collaboration, which observes events consistent with the unresolved
$P_c(4440) \! \to \! J/\psi \, p$ structure at
3.2$\sigma$~\cite{D0:2019sjy}.

Much about the pentaquark states remains unknown, starting with their
$J^P$ quantum numbers.  Indeed, their parity eigenvalues alone would
already reveal a great deal about their structure, since the parity
of an $S$-wave $J/\psi \, p$ (or $J/\psi \, \Lambda$) system is $-1$.
The possibility of molecules composed of weakly bound
$\Sigma^{(*)}_c$-${\bar D}^{(*)}$ pairs (also in an $S$ wave) was
examined even before the first LHCb paper~\cite{Wu:2010jy,
Wang:2011rga,Yang:2011wz,Karliner:2015ina}.  The proximity of the
$\Sigma_c \, \bar D^*$ threshold to $m_{P_c(4440)}$ and
$m_{P_c(4457)}$, the $\Sigma_c \bar D$ threshold to $m_{P_c(4312)}$,
and the $\Xi_c {\bar D}^*$ threshold to $m_{P_{cs}(4459)}$, has been
noted prominently in many publications, not least of which in LHCb's
own papers~\cite{Aaij:2019vzc,Aaij:2020gdg}.  The relevant thresholds
and $J^P$ values for these and related $S$-wave molecular states are
conveniently tabulated in  Ref.~\cite{Dong:2021bvy}.\footnote{It
should be noted that molecules in higher partial waves ($P,D,
\ldots$) are not logically precluded from forming, although the
presence of a centrifugal potential barrier for $L \! > \! 0$ may
interfere with the binding.  In addition, it seems likely that if
$P$-wave molecules exist, then the lower-energy (and more prominent)
$S$-wave molecules would be expected to be observed first.}  In
particular, no $P \! = \! +$ hidden-charm pentaquark molecular
thresholds occur below 4700~MeV; thus, if any of the known $P_c$
states is found to have $P \! = \! +$, it is not easily understood as
a hadronic molecule.  Since the structure $P_c(4380)$ is found to
carry opposite parity to the original
$P_c(4450)$~\cite{Aaij:2015tga}, then if it persists as a state, one
must conclude that at least one of the $P_c$'s is not
molecular.\footnote{Other issues with molecular interpretations of
$P_c$ involve the $J^P$ of the particle responsible for binding.  In
the case of $P_c(4312)$, a $0^-$ meson like $D$ cannot support a
trilinear coupling to another $0^-$ meson like $\pi$ or $\eta$; at
minimum, a $\rho$-like exchange would be necessary.  And if
nonstandard meson coupling fields like $\rho$ are sufficient for
binding hadronic molecules, then one might expect states, bound by
other light mesons, that include the isoscalar $\Lambda_c$ to have
appeared in the existing data.}

A large body of literature has examined the $P_c$ states using a
number of approaches: not only as hadronic molecules, but also
through quark-potential models, diquark models, QCD sum rules, and
others; for a discussion and numerous references, see the previously
cited reviews~\cite{Lebed:2016hpi,Chen:2016qju,Hosaka:2016pey,
Esposito:2016noz,Guo:2017jvc,Ali:2017jda,Olsen:2017bmm,
Karliner:2017qhf,Yuan:2018inv,Liu:2019zoy,Brambilla:2019esw}.  Most
of the relevant QCD sum-rule calculations~\cite{Wang:2020eep,
Wang:2019got,Azizi:2021utt,Azizi:2020ogm,Azizi:2016dhy}, but not all
(note Ref.~\cite{Chen:2015moa}), postdate these reviews.

In this work we apply a different approach, that of the dynamical
diquark model, to the quantitative study of the spectroscopic fine
structure of hidden-heavy-flavor pentaquarks.  The dynamical diquark
model is based upon the idea that multiquark systems occasionally
form configurations in which the attraction of two quarks in a
color-triplet channel is greater than that of either quark to the
nearest antiquark in a color-singlet channel.  One may then describe
the full hadron in terms of compounds of diquark [$\de \! \equiv \!
(Qq)_{\bar {\bf 3}}$]~\cite{Brodsky:2014xia} and triquark [$\bt \!
\equiv \! (\bar Q_{\bar {\bf 3}}
(q_1 q_2)_{\bar {\bf 3}})_{\bf 3}$]~\cite{Lebed:2015tna}
quasiparticles.  In order for this organization to be sensible, the
quasiparticles must achieve sufficient spatial separation to be
described as interacting through a potential $V(r)$.  If each
color-triplet quasiparticle component carries at least one heavy
quark (labeled above as $Q, \bar Q$), then $V(r)$ may be modeled
using the same static potentials as the ones that are calculated in
lattice simulations of quarkonium and its hybrid excitations, leading
to a description of exotic hadrons in terms of the Born-Oppenheimer
(BO) approximation.  This dynamical diquark model and the
spectroscopy of tetraquark and pentaquark states obtained from it
were first described in Ref.~\cite{Lebed:2017min}.

With a predictive model in hand, it becomes possible to study the
multiplet band structure for $c\bar c q\bar q^\prime$ and
$c\bar c qqq$ states numerically, as was first done in
Ref.~\cite{Giron:2019bcs}.  The fine structure of the ground-state
($S$-wave) and first excited-state ($P$-wave) tetraquark multiplets
were first studied in Refs.~\cite{Giron:2019cfc} and
\cite{Giron:2020fvd}, respectively.  The application to
$b\bar b q\bar q^\prime$ and $c\bar c s\bar s$ states appears in
Ref.~\cite{Giron:2020qpb}, $c\bar c c\bar c$ states in
Ref.~\cite{Giron:2020wpx}, and $c\bar c q\bar s$ states in
Ref.~\cite{Giron:2021sla}.

However, the fine structure of the hidden-charm pentaquark states has
not yet been analyzed in this model, largely due to the absence of
any particular $P_c$ state for which the $J^P$ quantum numbers are
definitively known.  In all the tetraquark cases, a specific state
[{\it e.g.}, $X(3872)$, $Z_b(10610)$, $X(4140)$] can be identified as
the cornerstone upon which the rest of the multiplet is built, but
the pentaquark sector to date lacks such a candidate.  However, one
expects that the same fine-structure Hamiltonian applies to the
tetraquark $\de$-$\bde$ and pentaquark $\de$-$\bt$ sectors, and
moreover, that comparing the nonstrange $c\bar c q\bar q^\prime$ and
strange $c\bar c q\bar s$ tetraquarks allows one to determine
properties of the $P_{cs}$ states from the corresponding
$P_c$ states~\cite{Giron:2021sla}.

The observation of the new $P_c(4337)$ state produces a very
interesting spectrum (see Table~\ref{tab:PentaMasses}): Two
narrow, closely spaced pairs of states [$P_c(4312) , P_c(4337)$ and
$P_c(4440) , P_c(4457)$] with nearly the same mass splitting.  The
dynamical diquark model, as shown in this paper, produces a unique
combination of such states in its pentaquark spectrum: The lower pair
are $J^P \! = \! {\frac{1}{2}}^+$, and the upper pair are $J^P \! =
\! {\frac{3}{2}}^+$.  Since all of these are $Q\bar Q qqq$ states
with $P \! = \! +$, they are $P$-wave states; we predict masses for
the other states in this multiplet, as well as those in the lower
$S$-wave multiplet.

We further use these results, as discussed above, to predict the
masses of the corresponding $P_{cs}$ states.  In this case, the
crucial ingredient of the analysis is a
comparison~\cite{Giron:2021sla} of the well-known hidden-charm
nonstrange states and the newly observed open-strange,
hidden-charm tetraquark states $Z_{cs}$:
\begin{eqnarray}
m_{Z_{cs}(3985)} & = & 3982.5^{+1.8}_{-2.6} \pm 2.1 \ {\rm MeV} ,
\nonumber \\
\Gamma_{Z_{cs}(3985)} & = & 12.8^{+5.3}_{-4.4} \pm 3.0 \ {\rm MeV} ,
\end{eqnarray}
from the BESIII Collaboration~\cite{Ablikim:2020hsk} in the process
$e^+ e^- \! \to \! K^+ (D_s^- D^{*0} \! + \! D_s^{*-} D^0)$, and
\begin{eqnarray}
m_{Z_{cs}(4000)} & = & 4003 \pm 6^{+ \ 4}_{-14} \ {\rm MeV} ,
\nonumber \\
\Gamma_{Z_{cs}(4000)} & = & \ 131 \pm 15 \pm 26 \ {\rm MeV} ,
\nonumber \\
m_{Z_{cs}(4220)} & = & 4216 \pm 24^{+43}_{-30} \ {\rm MeV} ,
\nonumber \\
\Gamma_{Z_{cs}(4220)} & = & \ 233 \pm 52^{+97}_{-73} \ {\rm MeV} ,
\end{eqnarray}
from the LHCb Collaboration in the process $B^+ \! \to \phi (J/\psi
\, K^+)$~\cite{Aaij:2021ivw}.

Other diquark-based pentaquark models for have appeared in the
literature.  For example, Ref.~\cite{Zhu:2015bba} was based upon the
diquark-triquark proposal of Ref.~\cite{Lebed:2015tna}, its model
containing spin-dependent, but not flavor-dependent, couplings.  The
model of Ref.~\cite{Ali:2019clg} is also of diquark-triquark type,
but both of the heavy quarks $c\bar c$ reside in the triquark.  The
very recent Ref.~\cite{Shi:2021wyt} appeared subsequent to the
discovery of $P_{cs}(4459)$, but is a diquark-diquark-antiquark
model.  This work is the first one to treat $P_c(4337)$ as a
diquark-based state.

This paper is organized as follows: In Sec.~\ref{sec:States}, we
reprise the notation of the model for identifying the pentaquark
states within their BO multiplets.  Section~\ref{sec:MassHam}
introduces the mass Hamiltonian for $S$-wave and $P$-wave multiplets,
and presents expressions for the masses of all their component
states.  In Sec.~\ref{sec:Analysis}, we analyze these expressions
numerically, using inputs from our previous work in the tetraquark
sector, and predict values for all unknown $S$- and $P$-wave
hidden-charm pentquarks in both light and open-strange sectors.
Section~\ref{sec:Concl} presents our conclusions.

\section{States of the Model}
\label{sec:States}

A cataloguing of  $\QQ q \hspace{1pt} q_1 q_2$ (and $\QQ q_1 \bar
q_2$) states in the dynamical diquark model, where
$q\! \in \!\{u,d,s\}$ and $q_i \! \in \! \{ u, d \}$, first appears
in Ref.~\cite{Lebed:2017min}.  The same notation, with small
modifications, is applied to $c\bar c s\bar s$ in
Ref.~\cite{Giron:2020qpb} and to $c\bar c c\bar c$ in
Ref.~\cite{Giron:2020wpx}.  All confirmed exotic candidates to date
have successfully been accommodated within the lowest ($\Sigma^+$)
Born-Oppenheimer (BO) potential of the gluon field connecting the
heavy diquark [$\de \! \equiv (Qq)$]-triquark
[$\bt \! \equiv \! (\bar Q (q_1 q_2))$] quasiparticles.  The
pentaquark BO potentials lack one discrete quantum number ($g,u$)
compared to those for $\de$-$\bde$ tetraquarks~\cite{Lebed:2017min}.
In all cases, $\de,\bt$ are assumed to transform as color triplets
(or antitriplets), and each quasiparticle contains no internal
orbital angular momentum.

In the case of $\QQ q \hspace{1pt} q_1 q_2$, the classification
scheme then begins with 3 possible core states in which the
$\de$-$\bt$ pair lie in a relative $S$ wave.  The most minimal form
of the model also restricts to states in which the diquark
$\dep \! = \! (q_1 q_2)$ internal to $\bt$, consisting of only light
$u,d$ quarks, is a ``good'' diquark: isosinglet and
$s_{\dep} \! = \! 0$, which is expected to be the most tightly bound
combination from light-hadron phenomenology~\cite{Jaffe:2004ph}
and from lattice simulations~\cite{Francis:2021vrr}.  In principle,
the ``bad'' (isotriplet, $s_{\dep} \! = \! 1$) combination could also
appear, but if so would form pentaquark states substantially higher
in mass.  Such states may be so broad as to have escaped detection at
LHCb\@.  Indicating the total spin $s$ of $\de, \bt$ by
$s_\de, s_\bt$, respectively, and including a subscript on the full
state to indicate its total spin, one obtains the spectrum
\begin{eqnarray}
J^{P} = \frac{1}{2}^{-}: & \ & P_{\frac{1}{2}} \equiv \left| \hf_\bt
, 0_\de \right>_{\frac{1}{2}} \,
, \ \ P_{\frac{1}{2}}^\prime \equiv \left| \hf_\bt , 1_\de
\right>_{\frac{1}{2}}\, ,
\nonumber \\
J^{P} = \frac{3}{2}^{-}: & \ & \pr \equiv \left| \frac{1}{2}_\bt ,
1_\de \right>_{\frac{3}{2}} \,
,
\label{eq:Swavediquark}
\end{eqnarray}
where again, the diquark $\dep$ spin $s_\dep$ is fixed to zero, and
so the triquark $\bt$ spin $s_{\bt} \! = \! \frac 1 2$.  Since 4
quark angular momenta ($s_q, s_\dep, s_Q, s_{\bar Q}$) are combined
here, one may transform these states into other convenient bases by
means of $9j$ angular momentum recoupling coefficients.  In
particular, in the basis of good total heavy-quark ($\QQ$) and
all-light baryonic ($\B \! \equiv \! q\dep \! = \! q \hspace{1pt} q_1
q_2$) spin, the transformation reads
\begin{eqnarray}
&&\left< (s_q \, s_{\dep}) s_\B , (s_Q \, s_{\bar Q}) s_\QQ
, S \, \right| \left. (s_q \, s_Q) s_\de , (s_{\bar q} \, s_{\bar Q})
s_\bt , S \right> \nonumber\\
&&=\left( [s_\B] [s_\QQ] [s_\de] [s_\bt] \right)^{1/2}
\begin{Bmatrix}
s_q & s_{\dep} & s_\B \\
s_Q & s_{\bar Q} & s_\QQ \\ 
s_\de & s_\bt & S
\end{Bmatrix}
\, , \ \ \label{eq:9jPenta}
\end{eqnarray}
using the abbreviation $[s] \! \equiv \! 2s + 1$ for the multiplicity
of a spin-$s$ state here and below.  Combining
Eqs.~(\ref{eq:Swavediquark}) and (\ref{eq:9jPenta}), one then obtains
\begin{eqnarray}
J^{P} = \frac{1}{2}^{-}: & \ & P_{\frac{1}{2}} = -\frac{1}{2} \left|
\frac{1}{2}_\B , 0_\QQ \right>_{\frac{1}{2}} + \frac{\sqrt{3}}{2}
\left| \frac{1}{2}_\B , 1_\QQ \right>_{\frac{1}{2}} \, ,
\nonumber \\
& & P_{\frac{1}{2}}^\prime = \frac{\sqrt{3}}{2} \left| \frac{1}{2}_\B
, 0_\QQ \right>_{\frac{1}{2}} + \frac{1}{2} \left| \frac{1}{2}_\B ,
1_\QQ  \right>_{\frac{1}{2}} \, , 
\nonumber \\
J^{P} = \frac{3}{2}^{-}: & \ & \pr = \left| \frac{1}{2}_\B , 1_\QQ
\right>_{\frac{3}{2}} \, .
\label{eq:SwaveQQ}
\end{eqnarray}
In this work, it is especially convenient to employ a basis of states
carrying a unique value of $s_\QQ$ and of $s_\B$ (the latter always
being $\frac 1 2$ since $s_{\dep} \! = \! 0$ and
$s_q \! = \! \frac 1 2$).  These states are the combinations
\begin{eqnarray}
\pt &\equiv& \ket*{\frac{1}{2}_\B,0_\QQ}_{\frac{1}{2}}=
-\frac{1}{2}P_{\frac{1}{2}}+\frac{\sqrt{3}}{2}P_{\frac{1}{2}}^\prime
\, , \nonumber\\
\ptp &\equiv& \ket*{\frac{1}{2}_\B,1_\QQ}_{\frac{1}{2}}=
+\frac{\sqrt{3}}{2}P_{\frac{1}{2}}+\frac{1}{2}P_{\frac{1}{2}}^\prime
\, , \nonumber\\
\pr & = & \left| \frac{1}{2}_\B , 1_\QQ \right>_{\frac{3}{2}} \, .
\label{eq:TildeStates}
\end{eqnarray}

The $\QQ q \hspace{1pt} q_1 q_2$ states in the multiplets
$\Sigma^+(1S)$ and $\Sigma^+(1P)$ are sufficient to accommodate all
particles considered in this work.  However, we note that
Ref.~\cite{Lebed:2017min} also provides a classification of higher
radial and orbital excitations [such as $\Sigma^+(nD)$], as well as
states in excited-glue BO potentials such as $\Pi^+$ (which are
exotic analogues to hybrid hadrons).  The corresponding
$\QQ q_1 \bar q_2$ states in the $\Sigma^+_g(1S)$ and
$\Sigma^+_g(1P)$ multiplets were first studied in
Refs.~\cite{Giron:2019cfc} and \cite{Giron:2020fvd}, respectively,
while radiative transitions between them are computed in
Ref.~\cite{Gens:2021wyf}.

\section{Pentaquark Mass Hamiltonian}
\label{sec:MassHam}


\subsection{$S$-Wave Hamiltonian and Mass Expressions}\label{swaves}



The $\Sigma^+(1S)$ (ground-state) pentaquark multiplet contains only
3 states, those listed in Eqs.~(\ref{eq:Swavediquark}) or
(\ref{eq:TildeStates}).  For this multiplet, the Hamiltonian closely
follows the form used for the $\Sigma^+_g(1S)$ tetraquark multiplet
in Ref.~\cite{Giron:2019cfc}, and reads
\begin{eqnarray}
H &=& M_0 \! +\Delta H_{\kqQ} + \Delta H_{\kdQ} + \Delta H_{\v8}
\nonumber \\
&=& M_0 +2\left[\kqQ\left(\vec{s}_q\cdot\vec{s}_Q\right)
+\kdQ\left(\vec{s}_\dep\cdot\vec{s}_{\bar{Q}}\right) \right] \nonumber \\
 &&+\v8\left(\lambda^8_a\vec{\sigma}_\dep\right)\cdot
 \left(\lambda^8_a\vec{\sigma}_q\right) \, .
\end{eqnarray}
Since the diquark $\dep$ is isoscalar by assumption, this Hamiltonian
lacks the nontrivial isospin dependence of its tetraquark analogue.
Under the assumption that $\dep$ maintains its flavor content
({\it i.e.}, does not exchange quarks with the other diquark), then
$\pi$-like and $K$-like exchanges do not occur.\footnote{Models
allowing for all such SU(3)$_{\rm flavor}$ exchanges do appear in the
literature~\cite{Shi:2021jyr,Shi:2021wyt}.}

Its matrix elements are 
\begin{eqnarray}
M&=& M_0 + \frac{\kqQ}{2}\left[2s_\de\left(s_\de+1\right)-3\right]
+\frac{\kdQ}{4}\left[4s_\bt\left(s_\bt+1\right)-3\right]\nonumber\\
&&+\frac{1}{6}\v8 C_1\left[4s_\B\left(s_\B +1\right)-3\right],
\end{eqnarray}
where 
\begin{eqnarray}
C_1 &=&
\begin{cases}
+2 & q_1,q_2,q \in \{u,d\},\\
-4 & q_1,q_2 \in \{u,d\} \cup q \in s.\\
\end{cases}
\end{eqnarray}
Because we consider only $s_\bt=s_\B=\frac{1}{2}$ (since $s_{\dep} \!
= \! 0$ in the dynamical diquark model), one can see that both the
triquark internal spin-spin coupling ($\Delta H_\kdQ$) and the
$\eta$-like coupling ($\Delta H_{\v8}$, first introduced in
Ref.~\cite{Giron:2021sla}), are zero.  Only the common mass
coefficient $M_0(1S)$ and the diquark $\de$ internal spin coupling
$\kappa_{qQ}$ survive.  In the order of the $s_{\QQ}$ eigenstates
$\pt, \ptp, \pr$ of Eqs.~(\ref{eq:TildeStates}), one finds
%
%
%
\begin{eqnarray}
\tilde{M}_{\frac{1}{2}^{-}} &=& M_0
\begin{pmatrix}
1 & 0 \\
0 & 1
\end{pmatrix}
+ \frac{1}{2}
\kqQ
\begin{pmatrix}
0 & \sqrt{3} \\
\sqrt{3} & -2
\end{pmatrix} \, ,\nonumber\\
M_{\frac{3}{2}^-} &=& M_0 +\frac{1}{2}\kqQ \, .
\end{eqnarray}
The diagonalized form of $M_{\frac 1 2}$, which also corresponds to
the basis of good $s_\de$ eigenvalues in the order given by
Eqs.~(\ref{eq:Swavediquark})
($P_{\frac{1}{2}}, P_{\frac{1}{2}}^\prime, P_{\frac{3}{2}}$), reads
\begin{eqnarray}
M_{\frac{1}{2}^{-}} &=&M_0
\begin{pmatrix}
1 & 0 \\
0 & 1
\end{pmatrix}
+ \frac 1 2 \kqQ
\begin{pmatrix}
-3 & 0 \\
0 & 1
\end{pmatrix} \, ,\nonumber\\
M_{\frac{3}{2}^-} &=& M_0 +\frac{1}{2}\kqQ \, .
\label{eq:SwaveMasses}
\end{eqnarray}
One finds (assuming $\kqQ \! > \! 0$) a unique $J^P \! = \!
{\frac 1 2}^-$ ground state and a degenerate pair with $J^P \! = \!
{\frac 1 2}^-$ and ${\frac 3 2}^-$.

\subsection{$P$-Wave Hamiltonian and Mass Expressions}\label{pwaves}

The $P$-wave states are obtained by combining those in the bases of 
Eqs.~(\ref{eq:Swavediquark}) or (\ref{eq:TildeStates}) with a unit
$L \! = \! 1$ of orbital angular momentum.  Using the latter basis
(unique $s_{\QQ}$ eigenvalues), one obtains the 7 states in the
multiplet $\Sigma^+(1P)$:
\begin{eqnarray}
J^{P}=\frac{1}{2}^{+}: & \ & \pt^{(L=1)}, \ \pt^{\prime (L=1)} , \
P_{\frac 3 2}^{(L=1)} \, , \nonumber \\
J^{P}=\frac{3}{2}^{+}: & \ & \pt^{(L=1)}, \ \pt^{\prime (L=1)} , \
P_{\frac 3 2}^{(L=1)} \, , \nonumber \\
J^{P}=\frac{5}{2}^{+}: & \ & P_{\frac 3 2}^{(L=1)} \, .
\label{eq:PwaveStates}
\end{eqnarray}
Table~\ref{tab:PwaveStates} collects the quantum numbers of these
states, including the total spin $s_B$ carried by the baryonic
combination $B$ of light quarks $(q \hspace{1pt} q_1 q_2)$ [which
always equals $\frac 1 2$ since $\dep \! \equiv \! (q_1 q_2)$ has
$s_{\dep} \! = \! 0$], and the total spin $S$ carried by all quarks.
In addition, we tabulate for completeness the relative amplitude
${\cal M}_{J_\B}$ within each state for each allowed eigenvalue
$J_{\B}$ of total angular momentum ${\bm J}_{\B} \! \equiv \! {\bm L}
+ {\bm s_B}$ carried by the light degrees of freedom.  These
recoupling amplitudes are given by
\begin{eqnarray}
\lefteqn{\mathcal{M}_{J_\B} \equiv \braket{
\left(L,s_\B\right), J_\B,s_\QQ,J}{L,\left(s_\B,s_\QQ\right),S,J}} &&
\nonumber\\
&=&(-1)^{L+s_\B+s_\QQ+J}\sqrt{\left[ J_\B \right] \left[S \right]}
\begin{Bmatrix}
L & s_\B & J_\B \\
s_\QQ & J & S
\end{Bmatrix} \, .
\end{eqnarray}
While a decomposition in $J_{\B}$ is not needed for the following
mass analysis, we anticipate its potential usefulness for computing
transition matrix elements, as is done for the tetraquark states in
Ref.~\cite{Gens:2021wyf}.

\begin{table}[t]
\caption{The 7 pentaquark $\QQ q \hspace{1pt} q_1 q_2$ states in the
$\Sigma^+(1P)$ multiplet of the dynamical diquark model, expressed in
the basis of good heavy-quark spin $s_{\QQ}$.  The model restriction
that the component diquark $\delta^{\prime} \! \equiv \! (q_1 q_2)$ carries zero
spin leads to the baryonic combination $B \! = \! (q \hspace{1pt} q_1
q_2)$ of the light quarks carrying $s_B \! = \! \frac 1 2$ for all
states.  $S$ is the total spin carried by quarks (the same as the
state subscript).  Also tabulated are the allowed values of light
degree-of-freedom total angular momentum $J_{B}$ and their amplitude
contribution $\mathcal{M}_{J_{B}}$ to the full state.}
\label{tab:PwaveStates}
\centering
\begin{tabular*}{\columnwidth}{@{\extracolsep{\fill}} c c c c c c c}\\
\hline\hline
State  & $J^{P}$ & $s_\B$ & $s_\QQ$ & $S$ & $J_\B$&
$\mathcal{M}_{J_\B}$ \\
\hline
$\pt^{(L=1)}$ & $\frac{1}{2}^{+}$ & $\frac{1}{2}$ & $0$ &
$\frac{1}{2}$ & $\frac{1}{2}$ & $+1$ \\
$\pt^{\prime (L=1)}$ & $\frac{1}{2}^{+}$ & $\frac{1}{2}$ & $1$ &
$\frac{1}{2}$ & $\frac{1}{2}$ & $-\frac{1}{3}$ \\
& & & & & $\frac{3}{2} $ & $+\frac{2\sqrt{2}}{3}$ \\
$P_{\frac 3 2}^{(L=1)}$ & $\frac{1}{2}^{+}$ & $\frac{1}{2}$ & $1$ &
$\frac{3}{2}$ & $\frac{1}{2}$ & $+\frac{2\sqrt{2}}{3}$ \\
& & & & & $\frac{3}{2} $ & $+\frac{1}{3}$ \\
$\pt^{(L=1)}$ & $\frac{3}{2}^{+}$ & $\frac{1}{2}$ & $0$ &
$\frac{1}{2}$ & $\frac{3}{2}$ & $+1$ \\
$\pt^{\prime (L=1)}$ & $\frac{3}{2}^{+}$ & $\frac{1}{2}$ & $1$ &
$\frac{1}{2}$ & $\frac{1}{2}$ & $-\frac{2}{3}$ \\
& & & & & $\frac{3}{2} $ & $+\frac{\sqrt{5}}{3}$ \\
$P_{\frac 3 2}^{(L=1)}$ & $\frac{3}{2}^{+}$ & $\frac{1}{2}$ & $1$ &
$\frac{3}{2}$ & $\frac{1}{2}$ & $+\frac{\sqrt{5}}{3}$ \\
& & & & & $\frac{3}{2} $ & $+\frac{2}{3}$ \\
$P_{\frac 3 2}^{(L=1)}$ & $\frac{5}{2}^{+}$ & $\frac{1}{2}$ & $1$ &
$\frac{3}{2}$ & $\frac{3}{2}$ & $+1$ \\
\hline\hline
\end{tabular*}
\end{table}
%


The minimal Hamiltonian for the first excited pentaquark multiplet,
$\Sigma^+(1P)$, closely follows the one used for the tetraquark
multiplet $\Sigma^+_g(1P)$ in Ref.~\cite{Giron:2020fvd}:
\begin{eqnarray}
H &=& M_0 +\Delta H_{\kqQ} + \Delta H_{\kdQ} + \Delta H_{\vls} +
\Delta H_{\vt} + \Delta H_{\v8} \nonumber\\
&=& M_0 +2\left[\kqQ\left(\vec{s}_q\cdot\vec{s}_Q\right) +
\kdQ\left(\vec{s}_\dep\cdot\vec{s}_{\bar{Q}}\right) \right]
\nonumber\\
&&+ \vls \vec{L}\cdot\vec{S} + \vt \hat{S}^{(\de \bt)}_{12}+
\v8\left(\lambda^8_a\vec{\sigma}_\dep\right)\cdot
\left(\lambda^8_a\vec{\sigma}_q\right) \, .
\label{eq:HamPwave}
\end{eqnarray}
In addition to a common mass [$M_0(1P)$ here], we allow for internal
spin-spin couplings for the diquark $\de$ ($\kqQ$) and the triquark
$\bt$ ($\kdQ$), as well as a spin-orbit term $V_{LS}$, a tensor term
$V_T$, and an $\eta$-like exchange term labeled by
$V_8$~\cite{Giron:2021sla}.  In addition, the numerical value of
$\kqQ$ in the $1P$ multiplet is expected to differ from that in
the $1S$ multiplet~\cite{Giron:2020fvd}.  The matrix elements of the
mass Hamiltonian read:
\begin{eqnarray}
M&=& M_0 + \frac{\kqQ}{2}\left[2s_\de\left(s_\de+1\right)-3\right]
+ \frac{\kdQ}{4}\left[4s_\bt\left(s_\bt+1\right)-3\right]\nonumber\\
&&+ \frac{\vls}{2}\left[J\left(J+1\right) - L\left(L+1\right)
- S\left(S+1\right)\right] \nonumber\\
&&+ \vt \langle S^{(\de\bt)}_{12} \rangle +\frac{1}{6}\v8 C_1
\left[4s_\B\left(s_\B +1\right)-3\right] \, .
\label{eq:MassPwave}
\end{eqnarray}

The tensor term labeled by $V_T$ differs somewhat from the primary
one studied in Ref.~\cite{Giron:2020fvd}, for which the spins
entering the operator are those of the individual light quarks
$q, \bar q$ within the diquarks.  Instead, the tensor operator used
here couples directly to the full quasiparticle $\de, \bt$ spins, and
thus is the analogue of the secondary possible tensor operator
studied in Ref.~\cite{Giron:2020fvd}, Appendix A.  Its matrix
elements are computed as
\begin{eqnarray}
&& \bra{L',S',J}\hat{S}^{(\de\bt)}_{12}\ket{L,S,J}
= (-1)^{S+J}\sqrt{30[L][L^\prime][S][S^\prime]} \nonumber \\
&&\times
\begin{Bmatrix}
J & S' & L' \\
2 & L & S
\end{Bmatrix}
\begin{pmatrix}
L' & 2 & L \\
0 & 0 & 0
\end{pmatrix}
\begin{Bmatrix}
s_{\bt} & s_\de & S \\
s^{\prime}_{\bt} & s^{\prime}_\delta & S'\\
1 & 1 & 2 
\end{Bmatrix}\nonumber\\
&&\times\bra{s^{\prime}_{\bt} \, } | \, \bm{\sigma}_\bt \, | \ket{\, s_{\bt}}
\bra{s^{\prime}_\delta \, } | \, \bm{\sigma}_\de \, | \ket{\, s_\de} \, .
\end{eqnarray}
Here, ${\bm \sigma}$ denotes not just spin-$\frac 1 2$ Pauli
matrices, but more generally twice the canonically normalized
generators ${\bf s}$ for arbitrary spin $s$, and the reduced matrix
elements of the angular momentum generators are given by
\begin{equation} \label{eq:Jreduced}
\left< j^\prime || \, {\bf j} \, || \, j \right> = \sqrt{j(2j+1)(j+1)}
\, \delta_{j^\prime j} \, .
\end{equation}
In particular,  if $s_\delta \! = \! 0$, then the reduced matrix
element (and hence $\langle \hat{S}^{(\de\bt)}_{12} \rangle$)
vanishes. 

Just as in the case of the $S$-wave states, one sees that both
$\Delta H_\kdQ \! = \! \Delta H_\v8 \! = \! 0$, and the surviving
matrix elements of Eq.~(\ref{eq:HamPwave}) for the $\Sigma^+(1P)$
multiplet, expressed in the same order as the states listed in
Eqs.~(\ref{eq:PwaveStates}) or Table~\ref{tab:PwaveStates}, then read



\begin{widetext}
\begin{eqnarray}
\tilde{M}_{\frac{1}{2}^{+}} &=& M_0
\begin{pmatrix}
1 & 0 & 0\\
0 & 1 & 0\\
0 & 0 & 1
\end{pmatrix}
+\frac{\kqQ}{2}
\begin{pmatrix}
-2 & \sqrt{3} & 0\\
\sqrt{3} & 0 & 0 \\
0 & 0 & 1
\end{pmatrix}
-\frac{\vls}{2}
\begin{pmatrix}
2 & 0 & 0\\
0 & 2 & 0\\
0 & 0 & 5
\end{pmatrix}
+\vt
\begin{pmatrix}
0 & 0 & \sqrt{6} \\
0 & 0 & \sqrt{2} \\
\sqrt{6} & \sqrt{2} & -4
\end{pmatrix} \, , \nonumber \\
\tilde{M}_{\frac{3}{2}^{+}} &=& M_0
\begin{pmatrix}
1 & 0 & 0\\
0 & 1 & 0\\
0 & 0 & 1
\end{pmatrix}
+\frac{\kqQ}{2}
\begin{pmatrix}
-2 & \sqrt{3} & 0\\
\sqrt{3} & 0 & 0 \\
0 & 0 & 1
\end{pmatrix}
+\frac{\vls}{2}
\begin{pmatrix}
1 & 0 & 0\\
0 & 1 & 0\\
0 & 0 & -2
\end{pmatrix}
+\frac{\vt}{5}
\begin{pmatrix}
0 & 0 & -\sqrt{15} \\
0 & 0 & -\sqrt{5} \\
-\sqrt{15} &  -\sqrt{5} & 16
\end{pmatrix} \, , \nonumber \\
M_{\frac{5}{2}^{+}}&=& M_0 +\frac{1}{2}\kqQ +\frac{3}{2} \vls
-\frac{4}{5}\vt \, .
\end{eqnarray}
\end{widetext}
These matrices, once diagonalized, provide the mass eigenvalues:
\begin{widetext}
\begin{eqnarray}
M_{\frac{1}{2}^+}&=& M_0
\begin{pmatrix}
1 & 0 & 0\\
0 & 1 & 0\\
0 & 0 & 1
\end{pmatrix}+\frac{\kqQ}{2}
\begin{pmatrix}
1 & 0 & 0\\
0 & -3 & 0\\
0 & 0 & 1
\end{pmatrix}
-\frac{1}{4}\vls
\begin{pmatrix}
7 & 0 & 0\\
0 & 4 & 0\\
0 & 0 & 7
\end{pmatrix}
-\vt
\begin{pmatrix}
2 & 0 & 0\\
0 & 0 & 0\\
0 & 0 & 2
\end{pmatrix}+\frac{\sqrt{3}}{4}\tilde{V}_1
\begin{pmatrix}
-1 & 0 & 0\\
0 & 0 & 0\\
0 & 0 & 1
\end{pmatrix} \, , \nonumber
\\
M_{\frac{3}{2}^+}&=&M_0
\begin{pmatrix}
1 & 0 & 0\\
0 & 1 & 0\\
0 & 0 & 1
\end{pmatrix}+\frac{\kqQ}{2}
\begin{pmatrix}
1 & 0 & 0\\
0 & -3 & 0\\
0 & 0 & 1
\end{pmatrix}-\frac{1}{4}\vls
\begin{pmatrix}
1 & 0 & 0\\
0 & -2 & 0\\
0 & 0 & 1
\end{pmatrix}
+\frac{8}{5}\vt
\begin{pmatrix}
1 & 0 & 0\\
0 & 0 & 0\\
0 & 0 & 1
\end{pmatrix}+\frac{\sqrt{3}}{20}\tilde{V}_2
\begin{pmatrix}
-1 & 0 & 0\\
0 & 0 & 0\\
0 & 0 & 1
\end{pmatrix} \, , \nonumber \\
M_{\frac{5}{2}^{+}}&=& M_0 +\frac{1}{2}\kqQ +\frac{3}{2} \vls
-\frac{4}{5}\vt \, ,
\label{eq:PwaveMasses}
\end{eqnarray}
\end{widetext}
where
\begin{eqnarray}
\tilde{V}_1&=&\sqrt{3\vls^2+16\vls\vt + 64\vt^2} \, , \nonumber \\
\tilde{V}_2&=&\sqrt{75\vls^2-320\vls\vt+448\vt^2} \, .
\label{eq:Vtildes}
\end{eqnarray}
The elements of these diagonalized matrices are presented in the
order of increasing mass eigenvalues in each $J^P$ sector, under the
assumption (as found for tetraquarks in Ref.~\cite{Giron:2020fvd})
that the contribution from $V_{LS}$ dominates the contribution from
$\kqQ$ (and from $V_T$).

Anticipating results from the analysis in the next section, we also
present the corresponding expressions to Eqs.~(\ref{eq:PwaveMasses})
when $V_T \! \to \! 0$.  Including up to linear order in $V_T$, one
finds
\begin{widetext}
\begin{eqnarray}
M_{\frac{1}{2}^+}&=& M_0
\begin{pmatrix}
1 & 0 & 0\\
0 & 1 & 0\\
0 & 0 & 1
\end{pmatrix}+\frac{\kqQ}{2}
\begin{pmatrix}
1 & 0 & 0\\
0 & -3 & 0\\
0 & 0 & 1
\end{pmatrix}
-\frac{1}{2}\vls
\begin{pmatrix}
5 & 0 & 0\\
0 & 2 & 0\\
0 & 0 & 2
\end{pmatrix}
-4\vt
\begin{pmatrix}
1 & 0 & 0\\
0 & 0 & 0\\
0 & 0 & 0
\end{pmatrix} \, , \nonumber
\\
M_{\frac{3}{2}^+}&=&M_0
\begin{pmatrix}
1 & 0 & 0\\
0 & 1 & 0\\
0 & 0 & 1
\end{pmatrix}+\frac{\kqQ}{2}
\begin{pmatrix}
1 & 0 & 0\\
0 & -3 & 0\\
0 & 0 & 1
\end{pmatrix}+\frac{1}{2}\vls
\begin{pmatrix}
-2 & 0 & 0\\
0 & 1 & 0\\
0 & 0 & 1
\end{pmatrix}
+\frac{16}{5}\vt
\begin{pmatrix}
1 & 0 & 0\\
0 & 0 & 0\\
0 & 0 & 0
\end{pmatrix} \, , \nonumber \\
M_{\frac{5}{2}^{+}}&=& M_0 +\frac{1}{2}\kqQ +\frac{3}{2} \vls
-\frac{4}{5}\vt \, .
\label{eq:PwaveMassesAPPROX}
\end{eqnarray}
\end{widetext}
It is worth emphasizing that the Hamiltonian of
Eq.~(\ref{eq:HamPwave}) with $V_T \! = \! 0$ is diagonal in the basis
of good $s_\de$ defined in Eqs.~(\ref{eq:Swavediquark}) [as is
apparent from Eq.~(\ref{eq:MassPwave}), since the $\kdQ$ and $V_8$
terms are also absent when $s_\dep \! = \! 0$].  The mass expressions
of Eqs.~(\ref{eq:PwaveMassesAPPROX}) with $V_T \! = \! 0$ therefore
refer to the diquark-spin basis of Eqs.~(\ref{eq:Swavediquark}),
specifically, in the order $P_{\frac{3}{2}}, P_{\frac{1}{2}},
P^\prime_{\frac{1}{2}}$.

\section{Analysis}
\label{sec:Analysis}

Using the results of the previous section, particularly
Eqs.~(\ref{eq:SwaveMasses}), (\ref{eq:PwaveMasses}), and
(\ref{eq:Vtildes}), we compute the mass eigenvalues for all
pentaquarks with flavor content $c\bar c q q_1 \hspace{-1pt} q_2$ and
$c\bar c uds$ in the ground-state multiplet [$\Sigma^+(1S)$] given in
Eqs.~(\ref{eq:Swavediquark}) or (\ref{eq:TildeStates}), and the first
excited-state multiplet [$\Sigma^+(1P)$] given in
Eqs.~(\ref{eq:PwaveStates}).  The measured masses of the 4 narrow
states [$P_c(4312)$, $P_c(4337)$, $P_c(4440)$, and $P_c(4457)$] in
Table~\ref{tab:PentaMasses} are used to assign these states to the
multiplet $\Sigma^+(1P)$, while we defer a discussion of the
problematic wide $P_c(4380)$ until later in this section.

The assignment of the known $P_c$ states to $\Sigma^+(1P)$ rather
than $\Sigma^+(1S)$ in the dynamical diquark model was first
noted in Ref.~\cite{Giron:2019bcs} to be much more natural, despite
the lack of clear experimental evidence for the lighter $S$-wave
states.  In that work, the argument rested upon the opposite-parity
nature of the (unresolved) $P_c(4450)$ and $P_c(4380)$.  Now, the
$\Sigma^+(1P)$ assignment can be made based upon the sheer
multiplicity of $P_c$ states: $\Sigma^+(1S)$ has only 3 states
[Eqs.~(\ref{eq:Swavediquark})], so that at least some of the $P_c$'s
must belong to a higher multiplet.  However, the model predicts a
$1P$-$1S$ mass splitting of $\sim \! 400$~MeV, rendering a mixed
$1P$-$1S$ assignment of the known $P_c$ states untenable.  In fact,
the closely (and almost equally) spaced pairs $P_c(4312)$-$P_c(4337)$
and $P_c(4440)$-$P_c(4457)$ have a completely natural identification
within $\Sigma^+(1P)$:  Anticipating our result that $V_T$ turns out
to be numerically small, and using the $V_T \! \to \! 0$ expressions
of Eqs.~(\ref{eq:PwaveMassesAPPROX}), one sees that the heaviest
${\frac{1}{2}}^+$ and lightest ${\frac{3}{2}}^+$ states are nearly
degenerate:
\begin{equation} \label{eq:Degen}
m^{(1)}_{{\frac{3}{2}}^+} - m^{(3)}_{{\frac{1}{2}}^+}
= \frac{36}{5} V_T \, ,
\end{equation}
and that, for $V_T \! = \! 0$, one finds two equally spaced pairs:
\begin{equation} \label{eq:ClosePair}
m^{(2)}_{{\frac{1}{2}}^+} - m^{(1)}_{{\frac{1}{2}}^+} =
m^{(2)}_{{\frac{3}{2}}^+} - m^{(1)}_{{\frac{3}{2}}^+} =
\frac{3}{2} V_{LS} - 2\kqQ \, .
\end{equation}
In contrast, the other 3 mass splittings of consecutive states in
$\Sigma^+(1P)$ equal $V_{LS}$ or $\frac{3}{2} V_{LS}$.  The analysis
of hidden-charm $P$-wave tetraquarks in Ref.~\cite{Giron:2020fvd}
produces values $V_{LS} \! \simeq \! 45$--60~MeV and
$\kqQc (1P) \! = \! 40$--45~MeV, so that typically,
$\left| \frac{3}{2} V_{LS} \! - \! 2\kqQc \right| \! \ll \! V_{LS}$.
Assuming that the corresponding hidden-charm nonstrange $P$-wave
pentaquarks produce similar numerical values for the coefficients
$V_{LS}$ and $\kqQc (1P)$ (as they must, since they arise from the
same dynamics), one expects the 7 masses in the $\Sigma^+(1P)$
spectrum to appear as two nearly degenerate values
[Eq.~(\ref{eq:Degen})], two closely spaced pairs
[Eq.~(\ref{eq:ClosePair})], and two heavier masses,
$m^{(3)}_{{\frac{3}{2}}^+}$ and $m_{{\frac{5}{2}}^+}$.

\subsection{The $c\bar c q q_1 \hspace{-1pt} q_2$ Sector, $P$ Wave}
\label{subsec:Analysis_ccqqqP}

Having thus identified specific candidates for the 4 known $P_c$
states:
\begin{eqnarray}
P_c(4312) & = & P_{\frac{1}{2}}^{(1)} \, , \nonumber \\
P_c(4337) & = & P_{\frac{1}{2}}^{(2)} \, , \nonumber \\
P_c(4440) & = & P_{\frac{3}{2}}^{(1)} \ {\rm and} \
P_{\frac{1}{2}}^{(3)} \, , \nonumber \\
P_c(4457) & = & P_{\frac{3}{2}}^{(2)} \, ,
\label{eq:PcAssign}
\end{eqnarray}
we first perform a least-squares fit to the $V_T \! = \! 0$ mass
expressions of Eqs.~(\ref{eq:PwaveMassesAPPROX}) and obtain values
for $M_0(1P)$, $V_{LS}$, and $\kqQc (1P)$.  We then predict the
masses for the remaining states of $\Sigma^+(1P)$.  The results are
presented in Table~\ref{tab:1PNonstrange}.

Since the $V_T \! = \! 0$ fit produces a figure of merit
$\chi^2 _{\rm min} \! < \! 1$, we conclude that the parameter $V_T$
is not actually needed for a complete description of the currently
available $P_c$ data.  Nevertheless, we perform a fit to the full
expressions of Eqs.~(\ref{eq:PwaveMasses})--(\ref{eq:Vtildes}), which
include $V_T$ nonlinearly.  Since 4 parameters are fit by 4 masses in
that case, the solution is unique, and we therefore do not propagate
the uncertainties on the parameters or masses for that fit.  These
results are also presented in Table~\ref{tab:1PNonstrange}.  We find
the essential result that $V_T$ is not merely statistically
unimportant (as shown by the $V_T \! = \! 0$ fit), but also that its
value in a fit with no free parameters is numerically small compared
to that of the other Hamiltonian parameters
[$M_0, V_{LS}, \kqQc(1P)$].  The most incisive test for the existence
of a nonzero $V_T$, as indicated by Eqs.~(\ref{eq:Degen}) and
(\ref{eq:PcAssign}), would be the resolution of $P_c(4440)$ into a
very closely spaced ${\frac{1}{2}}^+ \! , {\frac{3}{2}}^+$ pair.
Indeed, the difference between the ${\frac{3}{2}}^+$ and
${\frac{1}{2}}^+$ sides of Eq.~(\ref{eq:ClosePair}), using
Eqs.~(\ref{eq:PwaveMasses})--(\ref{eq:Vtildes}), is
$O(V_T^2/V_{LS})$; in comparison, its experimental value using the
masses in Table~\ref{tab:PentaMasses} is $-8.1 \! \pm \! 11.9$~MeV\@.

The hidden-charm $P$-wave pentaquarks studied here share several
similarities with the hidden-charm $P$-wave tetraquarks studied in
Ref.~\cite{Giron:2020fvd} within the dynamical diquark model, but
also feature some significant differences.  Both are modeled as heavy
(and therefore effectively semi-static) color-triplet quasiparticles
connected by the same orbitally excited ($L \! = \! 1$) color flux
tube, whose excitation energies are computed as BO potentials
obtained from specific lattice simulations, labeled here as JKM
(Refs.~\cite{Juge:2002br,Morningstar:2019}) and CPRRW
(Ref.~\cite{Capitani:2018rox}).  In both pentaquark and tetraquark
cases, the known states occupy the $1P$ levels of the ground-state BO
potential $\Sigma^+$, but since each tetraquark consists of a
[$\de \! = \! (cq)$]-[$\bde \! = \! (\bar c \bar q)$] pair, the
tetraquark BO potential is labeled by an additional $CP$ eigenvalue:
$\Sigma^+_g$.  In addition, the nonstrange pentaquark states in this
analysis all have $I \! = \! \frac 1 2$ because the diquark $\dep$
internal to the triquark $\bt$ is assumed to be isoscalar, so that
the overall isospin of the state is carried by the light quark in
$\de$.  The tetraquarks, in contrast, have isospin dependence via
interactions between the light quarks in the $\de$-$\bde$ pair (via
both spin-spin and tensor terms).

The two systems also feature some of the same operators in their
Hamiltonians, specifically the internal diquark spin coupling
$\kqQc (1P)$ and the spin-orbit coupling $V_{LS}$.  The analysis of
the hidden-charm $P$-wave tetraquarks is challenging because the 4
predicted $1^{--}$ states in $\Sigma^+_g(1P)$ can be assigned to
observed states [{\it e.g.}, $Y(4220)$] in a variety of
ways~\cite{Giron:2020fvd}, while as noted above, the analysis of the
hidden-charm $P$-wave pentaquarks is challenging due to a lack of
measured $J^P$ quantum numbers for any of the states.  Nevertheless,
the numerical values of $V_{LS}$ and $\kqQc (1P)$ for the two cases
are quite comparable: $V_{LS} \! \simeq \! 45$-60~MeV and
$\kqQc (1P) \! = \! 40$--45~MeV for the most plausible fits in
Ref.~\cite{Giron:2020fvd}, as compared with
$V_{LS} \! \simeq \! 80$~MeV and $\kqQc (1P) \! \simeq \! 50$~MeV
from Table~\ref{tab:1PNonstrange}.

Predictions for the masses of all 7 states in the $\Sigma^+(1P)$
multiplet are presented in Table~\ref{tab:1PNonstrange}.  $P_c(4312)$
is seen to be the lightest state of the multiplet, $P_c(4337)$ is the
next lightest, followed by the nearly degenerate pair coinciding with
$P_c(4440)$, and then $P_c(4457)$.  Notably, no lower or intermediate
states appear.  Finally, the heaviest ${\frac{3}{2}}^+$ and the
${\frac{5}{2}}^+$ lie much higher in mass; LHCb does present
$J/\psi \, p$ data from $\Lambda_b$ decays up to almost
5~GeV~\cite{Aaij:2019vzc}, but the statistics appear insufficient to
resolve states beyond about 4500~MeV\@.

Notably, no signal for $P_c(4337)$ is apparent in the LHCb
$\Lambda_b$ decay data, but only in their $B_s^0$ decay
data~\cite{LHCb:2021chn}.  On the other hand,
Ref.~\cite{LHCb:2021chn} sees no signal for $P_c(4312)$, and limited
phase space precludes observation of the higher $P_c$ states.  These
curious results from the same facility, taken at face value, suggest
different internal wave-function structures for $P_c(4312)$ and
$P_c(4337)$ being accessed through different processes.  Recalling
that the $V_T \! = \! 0$ mass eigenstates are those of good diquark
spin $s_\de$ [Eqs.~(\ref{eq:Swavediquark})], the LHCb data can be
explained if $\Lambda_b$ decays preferentially couple to states with
$s_\de \! = \! 1$ [and hence to $P_{\frac{3}{2}}$, the core quark
state of $P_c(4312)$], while $B_s^0$ decays preferentially couple to
states with $s_\de \! = \! 0$ [and hence to $P_{\frac{1}{2}}$, the
core quark state of $P_c(4337)$].  While the dynamical explanations
for these couplings are not immediately clear, one may observe that
the decay $B_s^0 \! \to \! J/\psi \, p \bar p$ requires the
annihilation of the initial valence quarks
$\bar b s \! \to \! \bar c c$ through the $t$-channel exchange of a
single virtual $W$ boson, while the initial $\Lambda_b$ light valence
quarks $ud$ in the dynamical diquark model persist through the decay
as the ``good'' diquark $\dep$.  Such differences could certainly
have a pronounced effect upon the internal spin structure of the
produced states.

Information can also be obtained from the quarkonium decays of the
states, assuming the conservation of heavy-quark spin $s_{\QQ}$.  The
underlying quark states
$P_{\frac{1}{2}}, P_{\frac{1}{2}}^\prime, P_{\frac{3}{2}}$ of
Eqs.~(\ref{eq:Swavediquark}), which we have found to coincide with
the mass eigenstates in the limit $V_T \! = \! 0$, are decomposed in
terms of $s_{\QQ}$ eigenvalues in Eqs.~(\ref{eq:SwaveQQ}).  Thus, for
example, $P_c(4457)$ coincides with $P_{\frac{3}{2}}$, which has only
an $s_{\QQ} \! = \! 1$ component.  Therefore, $P_c(4457)$ should (and
does) decay prominently to $J/\psi \, p$, but not to $\eta_c \, p$.

Lastly, we consider the troublesome $P_c(4380)$ opposite-parity
signal.  With the narrow $P_c$ states filling the positive-parity
$\Sigma^+(1P)$ multiplet, the very broad $P_c(4380)$ presumably
belongs to a negative-parity $S$-wave multiplet.  However,
calculations of multiplet-average masses in the dynamical diquark
model~\cite{Giron:2019bcs} predict the $\Sigma^+(1S)$ masses to be
nearly 400~MeV lower and the $\Sigma^+(2S)$ masses to be nearly
200~MeV higher than those in $\Sigma^+(1P)$.  We confirm these
results in this latest analysis, incorporating the best determination
of the $(cq)$ diquark mass from the latest analysis of
$c\bar c q^\prime \! \bar q$ tetraquarks in
Ref.~\cite{Giron:2021sla},\footnote{These values are quite comparable
to those obtained from other determinations, such as
$m_{\de(cq)} \! = \! 1975$~MeV from a constituent-quark
approach~\cite{Maiani:2004vq} or $1860 \! \pm \! 50$~MeV from QCD sum
rules~\cite{Kleiv:2013dta}.}
\begin{eqnarray} \label{eq:cqDiquarkV8}
m_{\de(cq)} & = & 1938.0 \pm 0.9 \ {\rm MeV} \ {\rm (JKM)} \, ,
\nonumber \\
& = & 1916.2 \pm 0.9 \ {\rm MeV} \ {\rm (CPRRW)} \, ,
\end{eqnarray}
to obtain the triquark masses $m_\bt$ presented in
Table~\ref{tab:1PNonstrange}.  If $P_c(4380)$ survives further
analysis as a distinct state, it cannot be of the same
diquark-triquark structure as the other $P_c$ states; the most likely
candidate would then be a $\bar D \Sigma^*_c$ threshold
effect or molecule~\cite{Shen:2016tzq}, which in its relative $S$
wave has the required negative parity.

\begin{table}
\centering
\caption{Calculations of Hamiltonian parameters and masses for the
$c\bar{c} q q_1 q_2$ $\Sigma^+(1P)$ states.  All
masses are in units of MeV\@.  The fit in the first column sets the
tensor coupling $V_T \! = \! 0$.  Boldface indicates best-fit masses
to measured values from Table~\ref{tab:PentaMasses}.
\label{tab:1PNonstrange}}
\begin{tabular}{c c c c c }\\
\hline\hline
$\chi^2_{\mathrm{min}}$ & \multicolumn{2}{c}{$0.468$} & \multicolumn{2}{c}{$0.000$}\\
$M_0(1P)$ & \multicolumn{2}{c}{$4495.4\pm 6.4$} &\multicolumn{2}{c} {$4492.0$}\\
$m_{\de = (cq)}$ & \multicolumn{2}{c}{$1927.1\pm 11.0$} & \multicolumn{2}{c}{$1927.1\pm 11.0$}\\
$m_\bt$ & \multicolumn{2}{c}{$2077.2\pm 11.6$} & \multicolumn{2}{c}{$2073.5\pm \ 9.0$} \\
$\kqQc (1P)$ & \multicolumn{2}{c}{$52.4\pm 6.3$} & \multicolumn{2}{c}{$49.9$}\\
$\vls$ & \multicolumn{2}{c}{$82.8\pm 5.6$} & \multicolumn{2}{c}{$80.2$}\\
$\vt$ & \multicolumn{2}{c}{$0.0$} & \multicolumn{2}{c}{$1.1$} \\
$M_{\frac{1}{2}^+}$ & \multicolumn{2}{c}{$\bm{4314.6}\pm 6.8$} & \multicolumn{2}{c}{$\bm{4311.9}$} \\
& \multicolumn{2}{c}{$\bm{4334.1}\pm 9.4$} & \multicolumn{2}{c}{$\bm{4337.0}$} \\
& \multicolumn{2}{c}{$4438.8 \ \pm 4.9$}  & \multicolumn{2}{c}{$4436.8$}  \\
$M_{\frac{3}{2}^+}$ & \multicolumn{2}{c}{$\bm{4438.8}\pm 4.9$} & \multicolumn{2}{c}{$\bm{4440.3}$}
\\
& \multicolumn{2}{c}{$\bm{4458.2}\pm 4.1$} & \multicolumn{2}{c}{$\bm{4457.3}$}  \\
& \multicolumn{2}{c}{$4563.0\pm 11.9$} & \multicolumn{2}{c}{$4557.3$} \\
$M_{\frac{5}{2}^+_{\vphantom{y}}}$ & \multicolumn{2}{c}{$4645.8\pm 17.3$} & \multicolumn{2}{c}{$4636.3$} \\
\hline\hline
\end{tabular}
\end{table}

\subsection{The $c\bar c q q_1 \hspace{-1pt} q_2$ Sector, $S$ Wave}
\label{subsec:Analysis_ccqqqS}

With specific diquark masses $m_\de$ [Eqs.~(\ref{eq:cqDiquarkV8})] 
and triquark masses $m_\bt$ (Table~\ref{tab:1PNonstrange}) in hand,
one can solve the coupled Schr\"{o}dinger equations in the dynamical
diquark model for all BO potentials, as is done in
Ref.~\cite{Giron:2019bcs}.  Here, we extract the value of $M_0(1S)$,
which is one of the two parameters appearing in the mass expressions
of Eqs.~(\ref{eq:SwaveMasses}), and present it in
Table~\ref{tab:1SNonstrange}.  The only additional parameter then
needed to predict all state masses in the $\Sigma^+(1S)$ multiplet is
$\kqQc (1S)$, which can be obtained from the latest hidden-charm
tetraquark analysis of Ref.~\cite{Giron:2021sla}:
\begin{equation} \label{eq:kqcS}
\kqQc (1S) = 25.6 \pm 5.0 \ {\rm MeV} \, .
\end{equation}
The masses of the 3 states of $\Sigma^+(1S)$ obtained from these
values are presented in Table~\ref{tab:1SNonstrange}.

Of course, no hidden-charm pentaquark states with such low masses
have yet been observed.  However, one notes two important points in
this regard: First, the region below about 4100~MeV in the data of
Ref.~\cite{Aaij:2019vzc} has rapidly vanishing $J/\psi \, p$ phase
space (threshold at 4035~MeV), meaning that the ground state at
$\simeq \! 4085$~MeV might be difficult to discern.  Even so, the
LHCb data from about 4100--4200~MeV appears as a broad enhancement
that could easily contain signals of resonances.  The $\Sigma^+(1S)$
states, carrying negative parity, are not protected by a centrifugal
barrier against $S$-wave fall-apart decays into $J/\psi \, p$, and
therefore can have substantial widths.  Second, the proposed
suppression of $\Lambda_b$ decays to $\Sigma^+(1P)$ states with
$s_\de \! = \! 0$, if it holds for the multiplet $\Sigma^+(1S)$,
eliminates the ground state $P_{\frac{1}{2}}$ in
Eqs.~(\ref{eq:Swavediquark}) from appearing in $\Lambda_b$ decays,
leaving only the (nearly degenerate) heavier
${\frac{1}{2}}^- \!$-${\frac{3}{2}}^-$ pair, $P_{\frac{1}{2}}^\prime$
and  $P_{\frac{3}{2}}$.  The LHCb data from $B_s^0$
decays~\cite{LHCb:2021chn} has substantially larger statistical
uncertainties than that from $\Lambda_b$ decays, but even so, hints
at some possible structures in the range of 4100--4200~MeV\@.

\begin{table}[ht]
\centering
\caption{Calculations of Hamiltonian parameters and masses for the
$c\bar{c}qq_1q_2$ $\Sigma^+(1S)$ states.  All masses are in units of
MeV\@.  The fit in the first (second) column uses values in which the
corresponding $M_0(1P)$ value (as well as $m_\de, m_\bt$) is obtained
from the first (second) column of Table~\ref{tab:1PNonstrange},
respectively.
\label{tab:1SNonstrange}}
\begin{tabular}{c c c c c}\\
\hline\hline
$M_0(1S)$ & \multicolumn{2}{c}{$4123.6\pm \ 1.7$} & \multicolumn{2}{c}{$4127.0\pm \ 1.7$}\\
$m_{\de = (cq)}$ & \multicolumn{2}{c}{$1927.1\pm 11.0$} & \multicolumn{2}{c}{$1927.1\pm 11.0$}\\
$m_\bt$ & \multicolumn{2}{c}{$2077.2\pm 11.6$} & \multicolumn{2}{c}{$2073.5\pm \ 9.0$} \\
$\kqQc (1S)$ & \multicolumn{2}{c}{$25.6 \pm 5.0$} & \multicolumn{2}{c}{$25.6 \pm 5.0$}\\
$M_{\frac{1}{2}^-}$ & \multicolumn{2}{c}{$4085.2\pm 7.7$} & \multicolumn{2}{c}{$4088.6\pm 7.7$} \\
& \multicolumn{2}{c}{$4136.4\pm 3.0$} & \multicolumn{2}{c}{$4139.8\pm 3.0$} \\
$M_{\frac{3}{2}^-_{\vphantom{y}}}$ & \multicolumn{2}{c}{$4136.4\pm 3.0$} & \multicolumn{2}{c}{$4139.8\pm 3.0$} \\
\hline\hline
\end{tabular}
\end{table}

\subsection{The $c\bar c uds$ Sector}
\label{subsec:Analysis_ccsqq}

Using the same triquark masses $m_\bt$ as in
Table~\ref{tab:1SNonstrange} and the $(cs)$ diquark mass
$m_{\de (cs)}$ obtained from the hidden-charm, open-strange
tetraquark states $Z_{cs}$ in Ref.~\cite{Giron:2021sla},
\begin{eqnarray} \label{eq:csDiquark}
m_{\de(cs)} & = & 2080.2 \pm 1.5 \ {\rm MeV} \ {\rm (JKM)} \, ,
\nonumber \\
& = & 2058.5 \pm 1.5 \ {\rm MeV} \ {\rm (CPRRW)} \, ,
\end{eqnarray}
one may immediately compute the $c\bar c uds$ multiplet-average
masses $M_0(1P)$ and $M_0(1S)$ for $\Sigma^+(1P)$ and $\Sigma^+(1S)$,
respectively, just as done in the previous subsections.  We also have
 the value of $\kqQs (1S)$ from Ref.~\cite{Giron:2021sla},
\begin{equation} \label{eq:kscS}
\kqQs (1S) = 109.8 \pm 1.1 \ {\rm MeV} \, ,
\end{equation}
which permits the immediate computation of the 3 $c\bar c uds$ states
in $\Sigma^+(1S)$.

However, we possess no independent determination of $\kqQs (1P)$,
since no $P$-wave $c\bar c q\bar s$ candidates have yet been
observed.  The large value of $\kqQs (1S)$ in Eq.~(\ref{eq:kscS}),
as compared to the value of $\kqQc (1S)$ in Eq.~(\ref{eq:kqcS}), is
argued in Ref.~\cite{Giron:2020fvd} to result from the heavier $s$
quark forming a $(cs)$ diquark that is more compact than $(cq)$,
which therefore subjects the $c,s$ quarks to much larger spin-spin
couplings.  Meanwhile, the larger value of $\kqQc (1P)$
($\simeq \! 50$~MeV from Table~\ref{tab:1PNonstrange}), as compared
to that of $\kqQc (1S)$ in Eq.~(\ref{eq:kqcS}), has been discussed in
Ref.~\cite{Giron:2020fvd} as a possible result of finite diquark
sizes leading to a sensitivity of their internal spin couplings on
the angular momentum of the flux tube connecting them.  One expects
this sensitivity to be greater for diquarks containing a light quark
$q$ than an $s$ quark.  Additionally, the $P$-wave Hamiltonian of
Eq.~(\ref{eq:HamPwave}) includes the spin-orbit coupling $V_{LS}$,
whose value might very well depend upon whether $(cs)$ or $(cq)$
diquarks are present.  For the analysis of the $\Sigma^+(1P)$
$c\bar c uds$ states, we simply adopt the numerical values of the
corresponding Hamiltonian couplings from the $\Sigma^+(1P)$
$c\bar c q q_1 \hspace{-1pt} q_2$ system given in
Table~\ref{tab:1PNonstrange}.  The results are presented in
Table~\ref{tab:1PStrange}.

One notes immediately from Table~\ref{tab:1PStrange} that the mass
predictions for the lightest two ${\frac 1 2}^+$ states are separated
by only about 20~MeV, and they bracket the measured mass of
$P_{cs}(4459)$ given in Table~\ref{tab:PentaMasses}.  These states
are the strange analogues of $P_c(4312)$ and $P_c(4337)$.  Indeed,
since the production of $P_{cs}(4459)$ via
$\Xi_b^- \! \to \! (J/\psi \, \Lambda) K^-$~\cite{Aaij:2020gdg} is a
strange analogue to the production of $P_c(4312)$ via
$\Lambda_b \! \to \! (J/\psi \, p) K^-$, one might expect that the
absence of $P_c(4337)$ in the latter predicts the absence of the
second-lightest ${\frac 1 2}^+$ $P_{cs}$ state in the former.  The
identification of $P_{cs}(4459)$ as the strange analogue of
$P_c(4312)$ in the dynamical diquark model was proposed in
Ref.~\cite{Giron:2021sla}.  Alternately, since the
$J/\psi \, \Lambda$ data does not yet have the same level of
statistics as the $J/\psi \, p$ data, the possibility of two closely
spaced ${\frac 1 2}^+$ states near 4459~MeV awaiting resolution is a
very real possibility.\footnote{A similar scenario is proposed in the
molecular picture~\cite{Du:2021bgb}, except in that case referring to
a ${\frac 1 2}^- \!$-${\frac 3 2}^-$ pair.}  The other $\Sigma^+(1P)$
$P_{cs}$ states in this calculation are over 100~MeV heavier, and
thus far only appear in the LHCb data~\cite{Aaij:2020gdg} as
tantalizing hints.

The predictions for the $\Sigma^+(1S)$ states, using
Eq.~(\ref{eq:kscS}), are presented in Table~\ref{tab:1SStrange}.
Again, the statistics of the LHCb data in Ref.~\cite{Aaij:2020gdg}
are not yet sufficient to determine whether significant $P_{cs}$
peaks reside in this mass range.  Similar comments to those for the
$\Sigma^+(1S)$ $P_c$ states apply in this case (noting that the
$J/\psi \, \Lambda$ threshold is 4213~MeV).

\begin{table}
\centering
\caption{Calculations of Hamiltonian parameters and masses for the
$c\bar{c}sq_1q_2$ $\Sigma^+(1P)$ states.  All masses are in units of
MeV\@. The fit in the first column sets the tensor coupling
$V_T \! = \! 0$.  The triquark mass $m_\bt$ values are obtained from
the corresponding column of Table~\ref{tab:1PNonstrange}.
\label{tab:1PStrange}}
\begin{tabular}{c c c c c}\\
\hline\hline
$M_0(1P)$ & \multicolumn{2}{c}{$4624.7 \pm \ 0.0$} & \multicolumn{2}{c}{$4621.3\pm \ 0.0$}\\
$m_{\de = (cs)}$ & \multicolumn{2}{c}{$2069.4\pm 10.9$} & \multicolumn{2}{c}{$2069.4\pm 10.9$}\\
$m_\bt$ & \multicolumn{2}{c}{$2077.2\pm 11.6$} & \multicolumn{2}{c}{$2073.5\pm \ 9.0$} \\
$\kqQc (1P)$ & \multicolumn{2}{c}{$52.4\pm 6.3$} & \multicolumn{2}{c}{$49.9$}\\
$\vls$ & \multicolumn{2}{c}{$82.8\pm 5.6$} & \multicolumn{2}{c}{$80.2$} \\
$\vt$ & \multicolumn{2}{c}{$0.0$} & \multicolumn{2}{c}{$1.1$}\\
$M_{\frac{1}{2}^+}$ &  \multicolumn{2}{c}{$4444.0\pm 14.4$} & \multicolumn{2}{c}{$4441.3$} \\
& \multicolumn{2}{c}{$4463.4\pm 11.0$} & \multicolumn{2}{c}{$4466.3$} \\
& \multicolumn{2}{c}{$4568.1\pm 6.4$} & \multicolumn{2}{c}{$4566.2$}\\
$M_{\frac{3}{2}^+}$ & \multicolumn{2}{c}{$4568.1\pm 6.4$} & \multicolumn{2}{c}{$4569.6$}
\\
& \multicolumn{2}{c}{$4587.6\pm 9.9$} & \multicolumn{2}{c}{$4586.6$}\\
& \multicolumn{2}{c}{$4692.3\pm 4.2$} & \multicolumn{2}{c}{$4686.4$}\\
$M_{\frac{5}{2}^+_{\vphantom{y}}}$ & \multicolumn{2}{c}{$4775.1\pm 9.0$} & \multicolumn{2}{c}{$4765.7$}\\
\hline\hline
\end{tabular}
\end{table}

\begin{table}
\centering
\caption{Calculations of Hamiltonian parameters and masses for the
$c\bar{c}sq_1q_2$ $\Sigma^+(1S)$ states.  All masses are in units of
MeV\@.   The fit in the first (second) column uses values in which the
corresponding $M_0(1P)$ value is obtained from the first (second)
column of Table~\ref{tab:1PStrange}, respectively.
\label{tab:1SStrange}}
\begin{tabular}{c c c c c}\\
\hline\hline
$M_0(1S)$ &  \multicolumn{2}{c}{$4258.6\pm 1.8$} & \multicolumn{2}{c}{$4255.2\pm 1.7$}\\
$\kappa_{sc} (1S)$ & \multicolumn{2}{c}{$109.8 \pm \ 1.1$} & \multicolumn{2}{c}{$109.8 \pm \ 1.1$}\\
$m_{\de = (cs)}$ &  \multicolumn{2}{c}{$2069.4\pm 10.9$} & \multicolumn{2}{c}{$2069.4\pm 10.9$}\\
$m_\bt$ & \multicolumn{2}{c}{$2077.2\pm 11.6$} & \multicolumn{2}{c}{$2073.5\pm \ 9.0$} \\
$M_{\frac{1}{2}^-}$ & \multicolumn{2}{c}{$4093.9\pm 2.4$} & \multicolumn{2}{c}{$4090.5\pm 2.3$}\\
& \multicolumn{2}{c}{$4313.5\pm 1.9$} & \multicolumn{2}{c}{$4310.1\pm 1.8$} \\
$M_{\frac{3}{2}^-_{\vphantom{y}}}$ & \multicolumn{2}{c}{$4313.5\pm 1.9$} & \multicolumn{2}{c}{$4310.1\pm 1.8$}\\
\hline\hline
\end{tabular}
\end{table}

\section{Conclusions}
\label{sec:Concl}

In this paper, we have shown that the dynamical diquark model
produces a spectrum of $c\bar c q q_1 \hspace{-1pt} q_2$ pentaquark
states that agrees very well with the 4 narrow $P_c$ resonances
observed by the LHCb Collaboration.  Their spectrum, as two closely
spaced pairs with nearly the same mass splitting, fits neatly with
the expected levels of the first excited-state multiplet
$\Sigma^+(1P)$ of the model, all members of which have positive
parity.  Specifically, $P_c(4312)$ and $P_c(4337)$ are predicted to
be ${\frac 1 2}^+$ states, $P_c(4440)$ is predicted to be a
yet-unresolved ${\frac 1 2}^+ \!$-${\frac 3 2}^+$ pair, and
$P_c(4457)$ is predicted to be ${\frac 3 2}^+$.  Moreover, the
numerical values of the Hamiltonian parameters (spin-spin,
spin-orbit) in this calculation are found to be closely comparable to
those obtained in a study of the negative-parity hidden-charm
tetraquark states [$Y(4220)$, {\it etc.}].

The 2 yet-unobserved states of the $\Sigma^+(1P)$ multiplet are
predicted to have much higher masses.  The broad, opposite-parity
$P_c(4380)$ structure does not fit into the model, and if it
persists, it is much more likely a threshold effect or molecule,
{\it e.g.}, caused by $\bar D \Sigma^*_c$.

The presence of $P_c(4312)$ and the absence of $P_c(4337)$ in one
production channel ($\Lambda_b$ decay), and vice versa in another
production channel ($B_s^0$ decay)---both seen at the same
experiment---is a remarkable feature.  Such a pattern can be
explained (if not yet fully understood) in this model by the fact
that the two states are composed of the same $[\bar c (ud)]$
color-triplet triquark, but have distinct internal spin structures
for their color-antitriplet diquark $(cq)$ component.  The
heavy-quark spin content of the states, as indicated, for example, by
the relative branching fractions of the resonances into $J/\psi$ or
$\eta_c$ final states, can also serve as a useful diagnostic in
uncovering the internal structure of the states.

Of exceptional importance in this analysis is the parity quantum
number for all of these states.  All $S$-wave molecules formed from a
ground-state charmed baryon ($\Sigma^+_c$ or $\Lambda_c$: parity~$+$)
and a ground-state charmed meson ($\bar D^{(*)}$: parity~$-$) have
negative parity.  If any of the narrow $P_c$ states is found to have
negative parity, then virtually all the analysis in this paper
becomes invalidated, or at least must be radically modified.

The 3 states of the ground-state multiplet $\Sigma^+(1S)$ in the
model, which do carry negative parity, are predicted to have masses
above the $J/\psi \, p$ threshold, but could have escaped detection by
LHCb either due to lower statistics in the near-threshold region, or
due to the states having large widths (as $S$-wave states, via their
$J/\psi \, p$ fall-apart modes).

The same model applied to the open-strange sector predicts
$P_{cs}(4459)$ to be the strange analogue to the ${\frac 1 2}^+$
$P_c(4312)$.  We also present predictions for all other $c\bar c uds$
states in the $\Sigma^+(1P)$ and $\Sigma^+(1S)$ multiplets, for which
the latest data gives only the vaguest indications.  Future
refinements of the statistics for channels like
$\Xi_b^- \! \to \! (J/\psi \, \Lambda) K^-$ will almost certainly
reveal the existence of further $P_{cs}$ resonances, and the spectrum
of these new states will provide crucial information for unraveling
their internal structure.

\vspace{1em}

\begin{acknowledgments}
This work was supported by the National Science Foundation (NSF) under 
Grant Nos.\ PHY-1803912 and PHY-2110278.
\end{acknowledgments}

\bibliographystyle{apsrev4-1}
\bibliography{diquark}
\end{document}